# A chip-scale polarization-spatial-momentum quantum SWAP gate in silicon nanophotonics

Xiang Cheng[1,*,+], Kai-Chi Chang[1,*,+], Zhenda Xie[1,2,3,*], Murat Can Sarihan[1], Yoo Seung Lee[1], Yongnan Li[1], XinAn Xu[2], Abhinav Kumar Vinod[1], Serdar Kocaman[2,4], Mingbin Yu[5,6], Patrick Guo-Qiang Lo[5,7], Dim-Lee Kwong[5], Jeffrey H. Shapiro[8], Franco N. C. Wong[8], and Chee Wei Wong[1,2,+]

[1] Fang Lu Mesoscopic Optics and Quantum Electronics Laboratory, University of California, Los Angeles, CA 90095, USA

[2] Optical Nanostructures Laboratory, Columbia University, New York, NY 10027, USA

[3] National Laboratory of Solid State Microstructures and School of Electronic Science and Engineering, Nanjing University, Nanjing 210093, China

[4] Department of Electrical Engineering, Middle East Technical University, 06800 Ankara, Turkey

[5] Institute of Microelectronics, Singapore 117685, Singapore

[6] State Key Laboratory of Functional Materials for Informatics, Shanghai Institute of Microsystem and Information Technology, and Shanghai Industrial Technology Research Institute, Shanghai 200050, China

[7] Advanced Micro Foundry, Singapore 117685, Singapore

[8] Research Laboratory of Electronics, Massachusetts Institute of Technology, Cambridge, MA 02139, USA

* these authors contributed equally to this work.

+ email: chengxiang@ucla.edu; uclakcchang@ucla.edu; cheewei.wong@ucla.edu

Recent progress in quantum computing and networking enables high-performance large-scale quantum processors by connecting different quantum modules. Optical quantum systems show advantages in both computing and communications, and integrated quantum photonics further increases the level of scaling and complexity. Here we demonstrate an efficient SWAP gate that deterministically swaps a photon's polarization qubit with its spatial-momentum qubit on a nanofabricated two-level silicon-photonics chip containing three cascaded gates. The on-chip SWAP gate is comprehensively characterized by tomographic measurements with high fidelity for both single-qubit and two-qubit operation. The coherence preservation of the SWAP gate process is verified by single-photon and two-

**photon quantum interference. The coherent reversible conversion of our SWAP gate facilitates a quantum interconnect between different photonic subsystems with different degrees of freedom, demonstrated by distributing four Bell states between two chips. We also elucidate the source of decoherence in the SWAP operation in pursuit of near-unity fidelity. Our deterministic SWAP gate in the silicon platform provides a pathway towards integrated quantum information processing for interconnected modular systems.**

Over the last few decades, quantum computing shows tremendous advancements towards realizing quantum supremacy [1,2] on different physical platforms [3,4]. Optical quantum systems appear to be the leading candidate for practical quantum computing with photons [5], which demonstrated to be scalable in principle with only linear optics [6]. Photons are potentially free from decoherence and easily manipulated in multiple degrees of freedom [7]. Nevertheless, effective interaction between photons is required to construct an optical quantum computer, which can be realized with optical quantum gates [8]. Universal quantum computing requires both single-qubit and two-qubit gate operations, the latter of which are usually probabilistic and cause large resource overhead scaling exponentially with the number of gates. Although cluster-state quantum computing has been proposed to drastically reduce resource overhead compared to the standard model [9, 10], these cluster states cannot be prepared deterministically and probabilistic quantum gates associated resource overhead remains high [11]. On the other hand, deterministic linear-optical quantum gates have been demonstrated utilizing several degrees of freedom (DoFs) of a single-photon for multiple-qubit encoding [12]. This scheme is equivalent to perform unitary operation on a multidimensional qubit (or qudit) encoded into single photon, thus can be implemented with linear optics [13]. Such deterministic two-qubit quantum gate will be beneficial for realizing a large-scale optical quantum computer due to its low resource overhead and high intrinsic success rate.

Optical quantum system also provides a natural integration of quantum computation and quantum communication, which is promising towards the ultimate goal of building a quantum internet [14, 15]. The quantum internet enables quantum communications among remote quantum devices via quantum links, which can significantly scale up the number of qubits for distributed quantum computing [16]. Photonics channels establish quantum links between distant stationary nodes with minimum loss and decoherence over long distance. Especially due to photons' well-defined Hilbert space in multiple DoFs, they are suitable to interconnect with different photonic platforms and increase the communications rates with high-dimensional encoding [17, 18]. Recently developed integrated quantum photonics opens another perspective for scaling up, taking advantage of wafer-scale fabrication process, a state-of-the-art large-scale quantum device with more than 550 optical components are demonstrated for multidimensional entanglement [19, 20]. These large-scale integrated photonics increases the scale and complexity of quantum circuits, and universal two-qubit unitary operation has been demonstrated exploiting high-dimensional entanglement in the path DoFs [21].

Exploiting the photonics platform with commercially available complementary metal–oxide–semiconductor (CMOS) compatible process, we demonstrate an efficient silicon SWAP gate that deterministically swaps the polarization qubit with spatial-momentum qubits from a single photon. Polarization DoF is easy to manipulate and measure using just waveplates and polarizing components, and spatial-momentum DoF is naturally compatible with integrated photonics for path-encoding and manipulation [19]. Our on-chip SWAP gate accesses these two DoFs by a concatenated scheme of three integrated controlled NOT gates: a specially designed momentum-controlled NOT (MC-NOT) gate sandwiched by two polarization-controlled NOT (PC-NOT) gates. We characterize the on-chip SWAP gate via state and process tomography with high-fidelity for both single-qubit and two-qubit operation. The preservation of quantum coherence in the on-chip SWAP operation of our silicon gate is verified via two-photon interference. The phase coherence of the on-chip SWAP operation is examined by single-photon self-interference with long-term stability. Furthermore, the reversible coherent conversion between polarization and spatial-momentum qubits of our on-chip SWAP gate enables quantum interconnect. We demonstrate the quantum photonic interconnectivity by distributing four Bell states between two SWAP gates with high state fidelity measured in the polarization DoF at the second chip. We also elucidate the source of possible errors for our silicon SWAP gate including imperfect qubit rotation, spatial-mode contamination, and unbalanced photon loss through theoretical model, and find good agreement with the measured truth table fidelity of the on-chip SWAP operation. Our chip-scale deterministic SWAP gate provides unitary operation in the control of single and entangled photons, and its coherent reversible conversion enables quantum photonic interconnect which will facilitate future distributed and cloud quantum computing [22-24].

**Results**

**SWAP gate configuration and chip implementation**

Figure 1a's left panel shows the logic circuit schematic of SWAP gate operation that swaps arbitrary values of qubits A and B without measuring or perturbing them. When qubits A and B are encoded, respectively, in the polarization and spatial-momentum mode of a single photon, SWAP gate operation can be realized with the three-gate cascade shown in Figure 1a's right panel [25]. In probabilistic linear optical quantum processing, most of the quantum logic operations are performed on two qubits, usually qubits of the same modality from two different photons. Here, a SWAP gate can coherently exchange states non-deterministically between qubits residing on different photons. In our single-photon two-qubit SWAP gate, qubit states are exchanged deterministically between the polarization and spatial-momentum degrees of freedom of the *same* photon, which suggests that robust on-chip multi-qubit single-photon logic of higher order should be sought [26-29].

The polarization qubit (*P*) is based on the two polarization eigenstates $|H\rangle$ and $|V\rangle$, which correspond to the transverse electric (TE) and transverse magnetic (TM) polarizations of our quantum photonic chip. Our experiments use a type-II phase matched spontaneous parametric downconversion (SPDC) waveguide source that produces $|V_S H_I\rangle$ biphotons in a single spatial mode, where the subscripts *S* and *I* denote the signal and idler qubits. The momentum qubit (*M*) is based on two spatial-momentum eigenstates $|T\rangle$ and $|B\rangle$, which correspond to the top and bottom channels of the quantum chip. Our experiments illuminate either the SWAP chip's top or bottom channels with polarization-rotated signal photons from the SPDC source, resulting in input state $|\Psi_T\rangle_{IN} = (|T_S H_S\rangle + e^{i\varphi}|T_S V_S\rangle)\otimes|H_I\rangle/\sqrt{2}$ for top-channel illumination and $|\Psi_B\rangle_{IN} = (|B_S H_S\rangle + e^{i\varphi}|B_S V_S\rangle)\otimes|H_I\rangle/\sqrt{2}$ for bottom-channel illumination, where $|H_I\rangle$ acts as a herald for the two qubits contained in its signal-photon companion. Defining $|0_{PS}\rangle = |H_S\rangle$, $|1_{PS}\rangle = |V_S\rangle$, $|0_{MS}\rangle = |T_S\rangle$, and $|1_{MS}\rangle = |B_S\rangle$ to be the logical-basis states, the input states become $|\Psi_T\rangle_{IN} = (|0_{MS}0_{PS}\rangle + e^{i\varphi}|0_{MS}1_{PS}\rangle)\otimes|H_I\rangle/\sqrt{2}$ and $|\Psi_B\rangle_{IN} = (|1_{MS}0_{PS}\rangle + e^{i\varphi}|1_{MS}1_{PS}\rangle)\otimes|H_I\rangle/\sqrt{2}$, which result in output states $|\Psi_T\rangle_{OUT} = (|1_{MS}1_{PS}\rangle + e^{i\varphi}|0_{MS}1_{PS}\rangle)\otimes|H_I\rangle/\sqrt{2}$ and $|\Psi_B\rangle_{OUT} = (|1_{MS}0_{PS}\rangle + e^{i\varphi}|0_{MS}0_{PS}\rangle)\otimes|H_I\rangle/\sqrt{2}$, respectively. The signal photon's polarization and spatial-momentum qubits have been swapped and undergone a bit flip.

The preceding SWAP operation is accomplished in our silicon photonics platform with three cascaded C-NOT gates designed so that the control and target qubits exchange roles in the middle C-NOT gate [30], as depicted in Figure 1a's right panel. In our architecture, the PC-NOT gate is realized by a silicon-photonics polarized directional coupler as shown in Figure 1b. The silicon MC-NOT gate is realized by a specially designed two-layer polarization structure that, as shown in Figure 1c, consists of two stages: (1) a polarization rotation stage, which tapers and rotates the qubit polarization by 90° and (2) a polarization-maintaining mode conversion stage, which converts the qubit mode profile to match the output waveguide. The polarization rotation stage is shown in Figure 1d. Because the polarization rotation and mode conversion are only implemented for the top channel, as shown in Figure 1f, the two-layer polarization structure thus performs the MC-NOT operation.

Each of the PC-NOT and MC-NOT gates has a silicon dioxide top cladding, with a rectangular silicon waveguide of 460 nm × 220 nm width-height cross-section, and with relatively small birefringence between the TE and TM modes for the polarization operations and diversity [31, 32]. The optimized PC-NOT gate has a waveguide-to-waveguide gap of 400 nm with a designed 11.5 µm coupling length, ensuring that the TE mode remains in its original waveguide while the TM mode crosses over to the other waveguide, with achieved average extinction ratio of ≈ 18 dB for different input-output ports and polarization combinations. The optimized MC-NOT gate has two 110 nm step-height layers, and with tapered widths down to 150 nm and uniform 30 nm lateral offsets. This specially designed polarization rotator requires two-level fabrication with two-mask alignment. The misalignment of the two masks (or levels) creates scattering losses and reduces the polarization extinction ratio [33], limiting the performance of the resulting MC-NOT gate and ultimately the SWAP gate. To overcome misalignment, we have developed a self-aligned two-level nanofabrication approach in order to achieve the high extinction ratio required for the polarization rotator. Two mask layers serve as the single mask for the first 220 nm Si reactive ion etch of the whole MC-NOT gate region, then the top layer is stripped via a resist developer, leaving the already-patterned hard mask. This hard mask is already self-aligned to the first etch, thus serves as the mask to define the 110 nm etch for the polarization rotator region in Figure 1c. Such self-alignment procedure eliminates the need for alignment between the two Si etch steps, and only two-level alignment before the first etch is needed for a relatively flat surface for lithography patterning, resulting in a guaranteed 30 nm layer-to-layer offset without alignment error. Our designed MC-NOT gate achieved a high extinction ratio of ≈ 20 dB for both TE and TM modes. The sidewall roughness is minimized for low waveguide loss, to ensure good SWAP gate performance. In addition, to ensure good coupling efficiency, adiabatic inverse tapers are designed for mode-index transformation at the input-output facets as shown in Figure 1e, with less than 3 dB loss for each facet.

The silicon PC-NOT and MC-NOT gates are individually characterized using a swept tunable laser (Santec TSL-510). The transmission spectra of both gates are measured via a free-space coupling system, which selects input and output channels for the gates. The input laser light's polarization is set by a polarizer and a half-wave plate for $|H\rangle$ or $|V\rangle$, and the output light is measured using a polarizer. The on-chip PC-NOT and MC-NOT gates are measured with more than 18 dB and 20 dB extinction ratio, respectively, over 100 nm span range in the C band. Next, we characterize the on-chip SWAP gate performance using the same coupling system for the four-basis states: $|TH\rangle$, $|TV\rangle$, $|BH\rangle$ and $|BV\rangle$. Consistent performance is achieved from 1550 nm to 1560 nm with extinction ratios of more than 12 dB. The cross-talk suppression of the SWAP gate is mainly bounded by the finite extinction ratios of the PC-NOT and MC-NOT gates, and the polarization misalignment between the output waveguide mode and the projection polarizers. The total insertion loss of the SWAP gate chip is estimated to be around 6 dB, which can be further reduced by better engineering of the coupler structure [34].

**Truth table of the on-chip SWAP gate**

With sufficiently low cross-talk measured between the basis states, we next examine the heralded single-photon two-qubit SWAP operation in the computational basis. Our single-photon two-qubit SWAP gate measurement setup is shown schematically in Figure 2a. Continuous-wave SPDC in a 1.5 cm ppKTP waveguide (AdvR) designed for type-II phase matching at ≈1556 nm produces orthogonally-polarized signal-idler biphotons [35]. The pump is a Fabry-Pérot laser diode stabilized with self-injection locking, through a double-pass first-order diffraction feedback with an external grating [36]. Tunable single-longitudinal mode lasing is achieved between 775.0 nm and 793.0 nm, enabling tunable SPDC with signal wavelengths from 1552.5 to 1559.6 nm, as shown in Figure 2a's inset. A long-pass filter blocks the residual pump photons after the SPDC, and an angle-mounted band-pass filter with 5 to 6 optical depth and a 95% passband transmission (Semrock NIR01-1570/3) further suppresses pump photons. Here, the biphoton state $|V_S H_I\rangle$ is generated by SPDC. The signal and idler photons are then separated by the polarization beam splitter. The signal photons are fed to the SWAP gate while the idler photons are directed to the superconducting nanowire single-photon detector (SNSPD; Photon Spot with ≈ 85% detection efficiency) for heralding. A two-in two-out free-space coupling system accesses the top and bottom channels of the SWAP chip at both its input and output facets. For each input channel, half-wave, quarter-wave and/or multi-order waveplates control the input polarization state for each measurement setup shown in Figure 2a. The polarization state of the signal photon becomes $|H\rangle$ or $|V\rangle$ or the superposition state given earlier according to the waveplate combination. The input spatial-momentum state is controlled by switching the input fiber (blue dashed line) to top or bottom channel of the SWAP gate, resulting in $|T\rangle$ or $|B\rangle$. For the truth table measurements, the input states to our SWAP gate are in the four-dimensional Hilbert space spanned by $|TH\rangle$, $|TV\rangle$, $|BH\rangle$ and $|BV\rangle$, corresponding to $|00\rangle$, $|01\rangle$, $|10\rangle$ and $|11\rangle$ in the logical basis. Polarizers P$_2$ and P$_3$ are rotated for polarization projection at the output ports. Coincidence counting is then performed using SNSPDs at the P$_2$ and P$_3$ outputs with internal timing delays to match that of the heralding detection. By recording the coincidence rates versus different polarization projections, we obtained the SWAP gate's truth table.

First, we measure the logical operation of our PC-NOT gate, by selecting an individual PC-NOT gate located on the same chip with our SWAP gate, with the same parameters as the SWAP gate's PC-NOTs. The characterization is performed using the measurement scheme shown in Figure 2a (I). Figure 2b shows the resulting measured truth table obtained for the four input states: $|00\rangle$, $|01\rangle$, $|10\rangle$ and $|11\rangle$ in the computational basis. The solid bars depict the experimentally measured truth table $M_{exp}$ while the transparent bars illustrate the ideal truth table $M_{ideal}$. The fidelity of the measured PC-NOT truth table with respect to the ideal one is calculated by $F = (1/4)Tr\left(\frac{M_{exp}M_{ideal}^T}{M_{ideal}M_{ideal}^T}\right)$. In our PC-NOT gate, we achieved an average fidelity of 97.8 $\pm$ 0.3 %., across the four basis states. We note that the residual deviation from unit fidelity is bounded by the PC-NOT's finite polarization-extinction ratio, and the $\approx$ 0.9 dB coupling difference between the $|H\rangle$ and $|V\rangle$ states. Similarly, to characterize our MC-NOT gate, we measure an individual polarization rotator located on the same chip, with the same parameters as the SWAP gate's MC-NOT. The test polarization rotator only has one spatial mode (top inset of Figure 2c), thus the truth table is only measured for two input polarization states: $|0\rangle$ and $|1\rangle$. The fidelity of the measured truth table in Figure 2c with respect to the ideal one is 98.0 $\pm$ 0.2 %. We can then infer the good performance of our on-chip MC-NOT gate, which is effectively a two-channel scheme of the polarization rotator and a silicon waveguide.

Having demonstrated good performance of each individual gate in the logical basis, we next measure the truth table of our on-chip SWAP gate. The truth table is measured by four measurements, each for four input states. We record a total of around 100,000 coincidence counts in 160 seconds for the truth table measurements yielding a truth table fidelity of 97.4 ± 0.2 % at 1557 nm, in support of the excellent performance in the logical basis. Truth table measurements are also performed at 1556nm and 1558 nm with similar fidelity, consistent with the broadband performance of the classical characterization. We attribute the deviations from unity in the truth-table fidelity mainly to the imperfect extinction ratio of the PC-NOT and MC-NOT gates and the MC-NOT gate's unbalanced photon loss.

**Quantum state and process tomographies for on-chip SWAP gate**

Although the truth table measures the two-qubit SWAP operation in the logical basis, quantum process tomography is required to completely characterize the Hilbert space of the SWAP gate operation [30, 37]. First, we use bulk optics to prepare the signal photons in an input set of six polarization states $\rho_{pol}$ ($|H\rangle$, $|V\rangle$, $|D\rangle$, $|A\rangle$, $|R\rangle$, $|L\rangle$) that are applied individually to the spatial input channels to the SWAP gate. Measuring the corresponding output spatial-momentum states $\rho_{sm}$ provides the quantum state tomography for these polarization inputs. The six input polarizations are shown in the center of Figure 3a as Bloch vectors. Two customized Mach-Zehnder interferometers (MZI) with over 20 dB extinction ratio and two tunable delay lines are used to adjust the input spatial-momentum modes for the on-chip SWAP gate and project the output qubit on a set of six spatial-momentum states after the SWAP operation, respectively, with the measurement setup shown in Figure 2a (II). The output spatial-momentum states are then analyzed to perform the quantum state tomography, with coincidence counts collected from the two output ports of MZI. The Bloch sphere representation of the measured output spatial-momentum states are shown in Figure 3a. The state fidelity is defined as: $F = (Tr(\sqrt{\sqrt{\rho_{pol}}\rho_{sm}\sqrt{\rho_{pol}}}))^2$, which describes the overlap between the input polarization states and the measured output spatial-momentum states. For different spatial inputs, we achieved an averaged fidelity $\bar{F}_{QST,T}$ of 97.2 ± 0.3% for $|T\rangle$ input, $\bar{F}_{QST,B}$ of 97.4 ± 0.3% for $|B\rangle$ input, $\bar{F}_{QST,+}$ of 97.1 ± 0.2% for $|+\rangle$ input and $\bar{F}_{QST,+i}$ of 97.0 ± 0.1% for $|+i\rangle$ input. These high-fidelity output spatial-momentum states, with an average fidelity of 97.3 ± 0.3%, confirm the successful single-qubit conversion from polarization qubit to spatial-momentum qubit.

Figure 3b shows the resulting process matrices of our SWAP gate for different spatial inputs.

This SWAP gate operation process can be represented by a reconstructed process matrix χ, defined as $\rho_{sm} = \sum_{mn} \chi E_m \rho_{pol} E_n^\dagger$, where $E_{m(n)}$ are the identity *I* and Pauli matrices *X*, *Y* and *Z* respectively. Thus, the SWAP gate's process matrix can be experimentally reconstructed by quantum state tomography measured in Figures 3a. The process fidelity is defined as $F_\chi = \frac{Tr(\chi\chi_i)}{Tr(\chi)Tr(\chi_i)}$, where $\chi_i$ is the theoretically ideal process matrix. The *X*, *Y* and *Z* component of the matrix χ represent the probability of a bit-flip or phase flip errors in the SWAP operation. We also evaluate the purity of the SWAP process matrix χ, defined as $P_\chi = \frac{Tr(\chi^2)}{Tr^2(\chi)}$, which is unity for an ideal process. Our SWAP gate is found to achieve a quantum process fidelity $\bar{F}_{\chi,T}$ of 95.5 ± 0.2% with process purity of 91.6 ± 0.2% for |T⟩ spatial-momentum mode input; $\bar{F}_{\chi,B}$ of 95.3 ± 0.2% with process purity of 91.6 ± 0.6% for |B⟩ input; $\bar{F}_{\chi,+}$ of 95.6 ± 0.2% with process purity of 91.5 ± 0.2% for |+⟩ input; and $\bar{F}_{\chi,+i}$ of 95.4 ± 0.1% with process purity of 91.2 ± 0.3% for |+i⟩ input. The average process fidelity for all spatial-momentum input modes is 95.5 ± 0.1%, verifying the single-qubit SWAP operation of our silicon gate from polarization to spatial momentum DoF.

For complete characterization of the two-qubit SWAP operation of our gate, we perform the full quantum process tomography. Additional to the process tomography measurement for single-qubit operation shown in Figure 2a (II), waveplates and polarizers are inserted before the MZI at the output of the chip for the polarization qubit analysis. First, we prepared 16 separable, linearly independent states $\rho_{sm,pol} = |i_{sm}j_{pol}\rangle$ as input two-qubit states, where $i_{sm} = 0, 1, +, +i$ and $j_{pol} = H, V, D, R$. The output states are projected in the same 16 state basis $\{|i_{sm}j_{pol}\rangle\}$. An averaged state fidelity of 96.1 ± 0.8% is achieved for the 16 input states. Then the process matrix χ is reconstructed similarly as for the single-qubit tomography using a block matrix of the measured density matrices as shown in Figure 3c [38]. We achieved a process fidelity of 94.9 ± 2.0% with a process purity of 93.3 ± 1.0%, which demonstrates the on-chip two-qubit SWAP operation of our gate. In addition, we note the sources of process fidelity non-ideality come from the bulk-optics imperfections in generating the input polarizations, the differential propagation loss and coupling efficiency mismatch between |H⟩ and |V⟩ states, and the residual misalignment of the spatial mode projection in the MZI.

**Quantum coherence of the on-chip SWAP operation**

An ideal SWAP operation is a coherent phase-preserving process. For a polarization input state of the form $|H\rangle + e^{i\varphi}|V\rangle$, the output state can be written as $|T\rangle + e^{i(\varphi+\delta)}|B\rangle$, where the phase difference φ between the orthogonal polarizations is transferred to the spatial modes and a constant phase δ accounts for the path length difference between the $|T\rangle$ and $|B\rangle$ spatial-momentum modes at the output. The $|T\rangle$ and $|B\rangle$ SWAP outputs of the signal photon are combined with a 50:50 fiber beam splitter, as shown in Figure 2a (III), whose outputs are detected in coincidence with the heralding idler photon to yield a self-interference measurement of the signal photon as a function of φ. An adjustable path delay *ΔT* (not illustrated in the schematic) is included in the bottom channel of the SWAP output for balancing the lengths of the two interferometer arms.

In our measurements we start with a 45° linearly polarized qubit $|D\rangle$, and the phase shift φ is introduced via a tuned pair of multi-order waveplates (illustrated in Figure 2a (III)) with their optic axes aligned to the $|V\rangle$ polarization. They are mounted on two motorized rotation stages for simultaneous counter-rotation along their optic axes. A tunable phase delay φ is imposed between $|H\rangle$ and $|V\rangle$ at the input by applying a rotation θ to one waveplate, while the transverse displacement of the beam is canceled with the counter rotation with the same angle magnitude for the other waveplate. The SWAP gate chip and the interference paths are carefully isolated from environmental noise for the phase-sensitive measurements. By sweeping the relative phase φ between the $|H\rangle$ and $|V\rangle$ polarizations of the input state of the signal photon, we probe the phase coherence of our SWAP gate operation by self-interference between the $|T\rangle$ and $|B\rangle$ output spatial-momentum states.

Figure 4a shows the self-interference fringes of the two spatial-momentum modes of the signal photon after the SWAP operation at different wavelengths. For the $|T\rangle$ spatial-momentum input state, a raw fringe visibility of 98.7 ± 0.2% is obtained (99.4% after background subtraction) at 1556 nm. This interference can also be observed when the polarization qubit is input through the bottom channel, with a raw visibility of 98.0 ± 0.2% (98.5% after background subtraction). The phase coherent polarization-to-spatial-momentum SWAP operation is also verified at 1557 nm and 1558 nm with high-visibility fringes as shown in Figure 4a, obtaining a wavelength-averaged single-photon self-interference visibility of 98.7% ± 0.4%. These observed high-visibility fringes demonstrate successful phase-coherence transfer from the input's polarization qubit to the output's spatial-momentum qubit. Moreover, we note that the phase interference is long-term robust and can maintain high visibility up to 96.6 ± 0.3% over 24 hours in free-running operation without feedback stabilization, verifying the phase-stable implementation of the on-chip single-photon two-qubit SWAP gate.

The coherence-preserved SWAP operation for two photons is further verified by off-chip Hong-Ou-Mandel (HOM) interference [39, 40], which measures the indistinguishability of the two photons over all DoFs. Implementing the experimental setup shown in Figure 2a (I), we fed both signal and idler photons to the on-chip SWAP gate via the 2-in-2-out coupling system, using HWPs to control the input polarization. At the output end, instead of the polarizers, we connect the two output channels to a HOM interferometer, consisting of a 50:50 fiber beamsplitter (FBS) and a delay line. A fiber polarization controller on one arm of the HOM interferometer ensures that the polarization of the two output photons will be the same at the FBS. By tuning the delay line, we can sweep the arrival time difference between the two output photons at the FBS and obtain the HOM interference dip. Figure 4b shows the measured HOM interference between the two output photons for different input polarization combinations. For $|T_S V_S\rangle \otimes |B_I H_I\rangle$ input, the HOM visibility of 96.9 (92.4) ± 1.4% is obtained after (before) background subtraction; for $|T_S H_S\rangle \otimes |B_I V_I\rangle$ input, the HOM visibility of 96.0 (91.0) ± 1.9% is achieved after (before) background subtraction. The slightly lower visibility for $|T_S H_S\rangle \otimes |B_I V_I\rangle$ input is because both signal and idler photons propagate through the polarization rotator on the upper arm of the SWAP gate, which introduces extra loss compared to the case for $|T_S V_S\rangle \otimes |B_I H_I\rangle$ input. The HOM dip width indicates the two-photon coherence time, which is measured to be 3.17 ± 0.02 ps for $|T_S V_S\rangle \otimes |B_I H_I\rangle$ input, and 3.11 ± 0.03 ps for $|T_S H_S\rangle \otimes |B_I V_I\rangle$ input. The indistinguishability of the SPDC photon pairs is also examined using the same HOM interferometer, with HOM visibility of 97.9 (93.4) ± 1.0% after (before) background subtraction and two-photon coherence time of 3.15 ± 0.02 ps. The small degradation of the HOM interference visibility and coherence time after the SWAP operation compared to the SPDC source unambiguously proves the preservation of the quantum coherence in the on-chip SWAP process. The observed HOM interference dip also verifies the indistinguishability between the two output spatial modes of the on-chip SWAP gate, which is crucial for path-mode entanglement generation on chip enabled by quantum interference [41, 42].

**Quantum photonic interconnect between two SWAP gate chips**

With coherence-preserved SWAP gate operation verified with high-fidelity on our silicon chip, we further demonstrate an efficient quantum photonic interconnect between different DoFs utilizing the reversible conversion of our on-chip SWAP process. The experimental scheme of the chip-to-chip interconnect is illustrated in Figure 4c. The input two-qubit maximally entangled state $\varphi$ is prepared in polarization basis. Polarization Bell state $|\Psi^+\rangle = (|HV\rangle + |VH\rangle)/\sqrt{2}$ is first generated by temporally overlapping the SPDC biphotons at a beamsplitter with orthogonal polarization [36]. The signal and idler photons are then fed to the $|T\rangle$ and $|B\rangle$ channels of the first SWAP gate chip, respectively. The input state thus can be written as: $\varphi = (|H_SV_I\rangle + |V_SH_I\rangle)\otimes|T_SB_I\rangle/\sqrt{2}$. The first SWAP gate then deterministically swaps the entanglement from polarization to spatial-momentum, yielding $\varphi_{sm} = (|T_SB_I\rangle + |B_ST_I\rangle)\otimes|H_SV_I\rangle/\sqrt{2}$. The output spatial-momentum entangled state is transmitted to the second SWAP gate chip via single-mode fiber link, where the polarization rotation during transmission is compensated by the QWPs and HWPs at the input of the second chip. The second SWAP gate has the same structural parameters as the first SWAP gate, and is characterized with truth table $\bar{F}_{gate,truth}$ of 97.2 ± 0.3%. The spatial-momentum entangled state is then reversibly converted to polarization entangled state $\varphi$ by the second SWAP gate, and measured by polarization analyzers consisting of a QWP, HWP and a polarizer to perform quantum state tomography. By adjusting the HWPs and QWPs at the input of the first SWAP gate, the other three Bell states $|\Psi^-\rangle$, $|\Phi^+\rangle$, and $|\Phi^-\rangle$ can be produced for chip-to-chip distribution [43].

Figure 4d shows the experimentally reconstructed density matrices for four polarization Bell states. The state fidelity is calculated by: $F_{Bell} = (Tr(\sqrt{\sqrt{\rho_{Ideal}}\rho_{Bell}\sqrt{\rho_{Ideal}}}))^2$, which describes the overlap between the ideal Bell states and the measured states. The fidelities of the reconstructed density matrices compared to the corresponding Bell states are: $F_{|\psi^+\rangle} = 92.5 \pm 0.3\%$, $F_{|\psi^-\rangle} = 90.4 \pm 0.5\%$, $F_{|\Phi^+\rangle} = 92.0 \pm 0.6\%$ and $F_{|\Phi^-\rangle} = 91.1 \pm 0.7\%$, with an averaged fidelity of 91.5 ± 0.8%. The nonideality of the fidelity is attributed to the waveguide loss, unbalanced coupling efficiency, imperfect rotation of the polarization elements and misalignment of the polarization analyzers. The chip-to-chip distribution of the four Bell states demonstrates the coherent reversible conversion of our SWAP gate between polarization and spatial-momentum DoF. This demonstration also provides a practical tool for quantum interconnect between distant photonic platforms with different DoFs towards distributed quantum computation and quantum sensing [17, 44].

**Discussion**

We have successfully demonstrated a deterministic single-photon two-qubit SWAP gate between polarization and spatial-momentum on silicon chip. The performance of our on-chip SWAP gate can be further improved by optimizing the fabrication parameters and chip coupling. We note that the deviations from unity in the truth-table fidelity mainly arise from the imperfect extinction ratio of the PC-NOT and MC-NOT gates and the MC-NOT gate's unbalanced photon loss. These non-idealities can be mitigated by more adiabatic polarization-mode conversion and tighter suppression of the cross-polarization. In addition, we note that the waveguide loss and unbalanced coupling efficiency between the $|H\rangle$ and $|V\rangle$ states contribute to the truth-table fidelity reduction by $\approx 0.5\%$. With recent progress on integrated polarization devices, polarization beamsplitters with over 35 dB extinction ratio and polarization rotators with low insertion loss have been realized on silicon platform [32], which can bring our chip's truth-table fidelity up to near unity. In addition, a silicon based MZI with over 66 dB extinction ratio was achieved, which will further improve the path-mode projection for quantum state tomography measurements [45].

The quantum coherence is preserved during the on-chip SWAP process and the coherent reversible conversion enables quantum interconnectivity between two chips. We note that conversion of photonic quantum states between different DoFs has been demonstrated on-chip [40, 46], however, none of which demonstrated an on-chip two-qubit SWAP gate operation. With the CMOS-compatible silicon chip-scale platform, high-density photonic integration involving different DoFs might be possible for future applications [47], extending to high-dimensional quantum gate operation [48, 49], with intrinsic good phase stability and compactness. The demonstrated quantum photonic interconnect can facilitate applications exploiting polarization and spatial-momentum entanglement between chip-based subsystems. In addition, the compatibility with microelectronics enables monolithic integration of photon sources, logic circuits, and detectors on silicon platform [20, 50]. Our on-chip SWAP gate paves the way for deterministic chip-scale quantum information processing and provides a photonic quantum interface for large-scale interconnected quantum information systems towards a quantum internet.

**Acknowledgements**

The authors acknowledge discussions with H. Liu, K. Yu, Y. Cho, A. Veitia, T. Zhong, F. Sun, and the scanning electron micrograph assistance from J. F. McMillan. This work is supported by the National Science Foundation (EFRI-ACQUIRE 1741707, QII-TAQS 1936375, 1919355 and 2008728) and the University of California National Laboratory research program (LFRP-17-477237).

**Author contributions**

X.C., K.C., Z.X., and Y.S.L. performed the measurements, Y.L., S.K., and X.X. performed the design layout, M.Y. P.G.Q.L and D.L.K. performed the device nanofabrication, X.C., M.C.S., X.X., A.K.V., J.H.S, and F.N.C.W. contributed in the theory and numerical modeling, and X.C., Z.X., X.X., J.H.S, F.N.C.W., and C.W.W. wrote the manuscript with contributions from all authors.

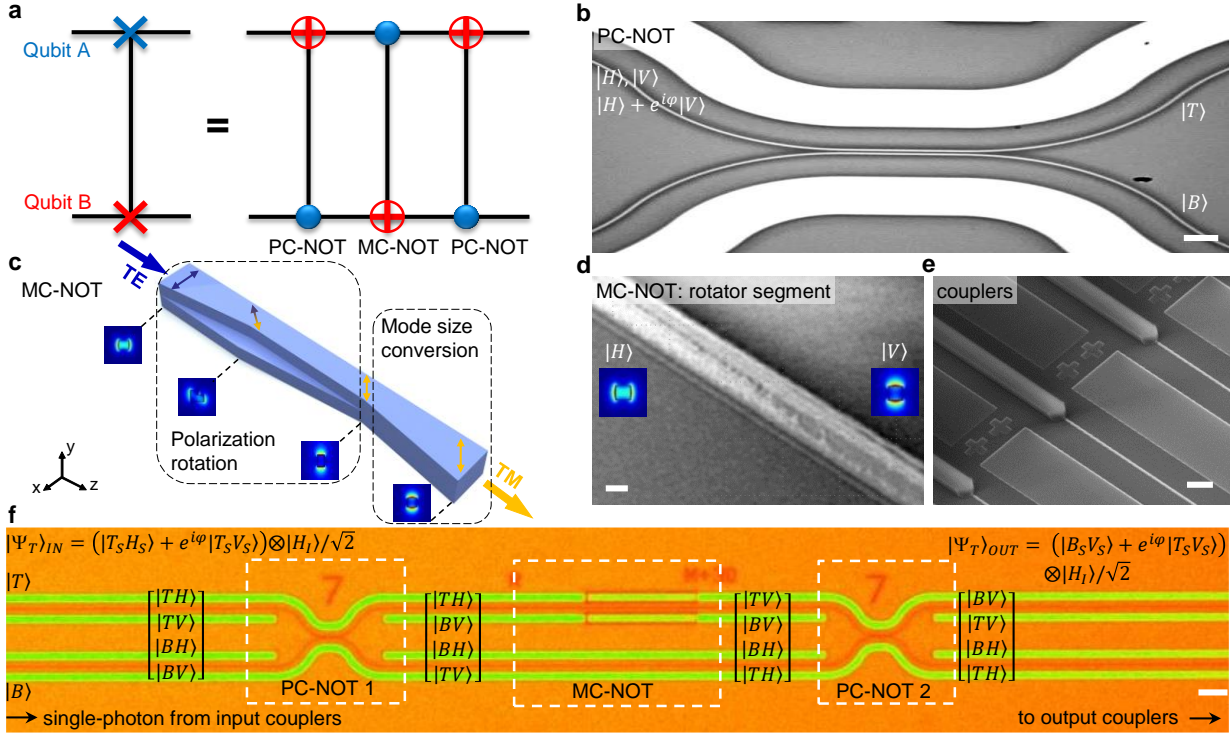

**Figure 1 | A chip-scale polarization-spatial single-photon two-qubit SWAP gate. a,** Illustrative logic circuit of the two-qubit SWAP gate. This can be realized, for single photon carrying qubit A in its polarization and qubit B in its spatial-momentum modes, by applying sequentially a PC-NOT gate, a MC-NOT gate and another PC-NOT gate, which are controlled by qubits A, B and A, respectively. **b,** Scanning electron micrograph (SEM) of the chip-scale SWAP gate's first-stage PC-NOT gate realized by an optimized integrated-photonics polarized coupler. Scale bar: 2 μm. **c,** Schematic of an integrated two-level polarization rotator with polarization rotation and mode size conversion sections, enabling the second-stage MC-NOT gate for the SWAP operation. **d,** SEM of the MC-NOT gate's nanofabricated polarization rotation segment. Scale bar: 500 nm. **e,** SEM of the inverse taper couplers for improved free-space qubit-to-chip coupling. Scale bar: 20 μm. **f,** Optical micrograph of the complete SWAP gate operation using the cascaded PC-NOT / MC-NOT / PC-NOT architecture. Scale bar: 10 μm. An example of input state ($|\Psi_T\rangle_{IN}$) for top channel of the SWAP gate is denoted, leading to the output state $|\Psi_T\rangle_{OUT}$, where signal photon's polarization qubit is swapped to spatial-momentum qubit. State vectors at each NOT gate segment represent the resulting states of each gate operation on four possible input states [$|TH\rangle, |TV\rangle, |BH\rangle, |BV\rangle$].

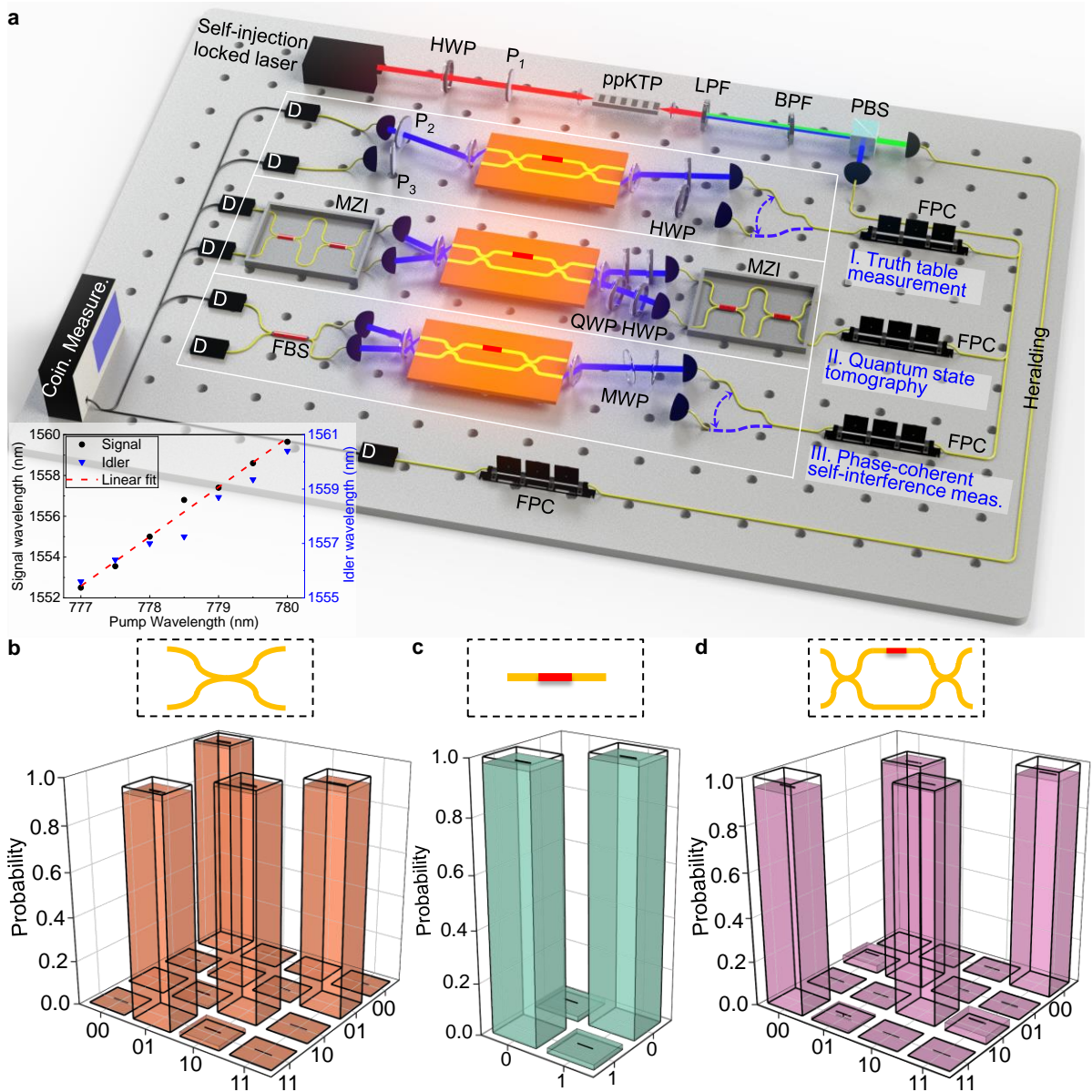

**Figure 2 | Experimental configuration for characterization of the single-photon two-qubit SWAP gate and truth table measurements. a,** Schematic of the heralded single-photon measurements generated via SPDC, with the three modular experiments: (I) truth table characterization, (II) quantum state tomography, and (III) phase-coherence self-interference measurements. Input polarization qubits are controlled by a half-wave plate (HWP) and then fed to the gate through free-space coupling. Output spatial-momentum qubits are examined by a polarization analyzer for truth table measurement, by Bloch state measurements for the quantum state tomography, and interfered in 50:50 beam splitter for phase preservation checks on the SWAP operation. Successful SWAP operation is heralded by coincidence counting between signal and

heralding channel. P$_i$: linear polarizer. LPF: long-pass filter. BPF: band-pass filter. PBS: polarization beam splitter. FPC: fiber polarization controller. MWP: multi-order waveplate. QWP: quarter-wave plate. MZI: Mach-Zehnder interferometer. FBS: 50:50 fiber beam splitter. D: superconducting nanowire single-photon detector. Bottom left inset: signal and idler photon wavelengths as a function of the pump wavelength. The red dashed line is a linear fit on the signal photon wavelength. **b-d,** Measured (solid bars) and ideal (transparent bars) truth table for the PC-NOT gate, MC-NOT gate and SWAP gate in the computational basis. A total of around 100,000 coincidence counts is recorded in 160 seconds for each measurement, yielding an average fidelity of 97.8 $\pm$ 0.3 %, 98.0 $\pm$ 0.2 % and 97.4 $\pm$ 0.2 %, respectively.

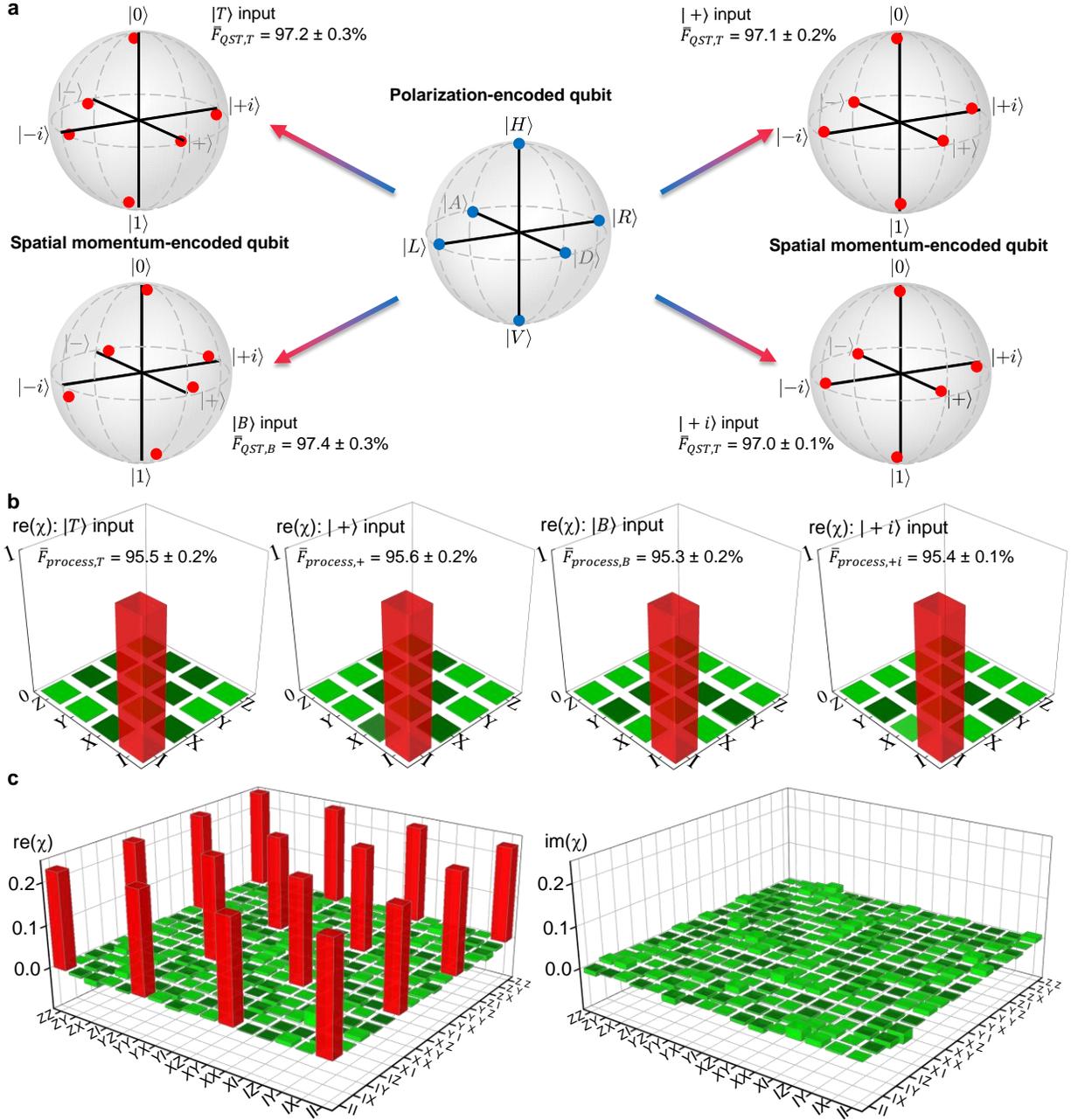

**Figure 3 | Quantum state and process tomographies for one-qubit and two-qubit SWAP operation. a,** Measured output spatial-momentum encoded states $|0\rangle$, $|1\rangle$, $|+\rangle$, $|-\rangle$, $|i\rangle$ and $|-i\rangle$ by Mach-Zehnder interferometer represented by red dots on the Bloch sphere for input polarization qubits prepared in $|T\rangle$, $|B\rangle$, $|+\rangle$ and $|+i\rangle$ spatial-momentum modes. Indicated fidelity represents the average over the six measured states. Middle: Bloch-sphere representation of six polarization-encoded input states $|H\rangle$, $|V\rangle$, $|D\rangle$, $|A\rangle$, $|R\rangle$ and $|L\rangle$ prepared by bulk optics (blue dots). **b,** Real parts of the reconstructed single-qubit process matrix χ of the SWAP gate for $|T\rangle$, $|B\rangle$, $|+\rangle$ and

$|+i\rangle$ spatial-momentum mode inputs, with an averaged process fidelity of 95.5 ± 0.1% and process purity of 91.5 ± 0.2%. All imaginary elements of the process matrix are smaller than 0.05. **c,** The reconstructed process matrix $\chi$ of the single-photon two-qubit SWAP gate. Additional to Figure 2a (II), HWPs, QWPs and polarizers are inserted before the MZI at the chip output for the polarization qubit analysis. Quantum state tomography results of the 16 input two-qubit states yield an averaged state fidelity of 96.1 ± 0.8%. The two-qubit SWAP process fidelity is measured to be 94.9 ± 2.0% with process purity of 93.3 ± 1.0%.

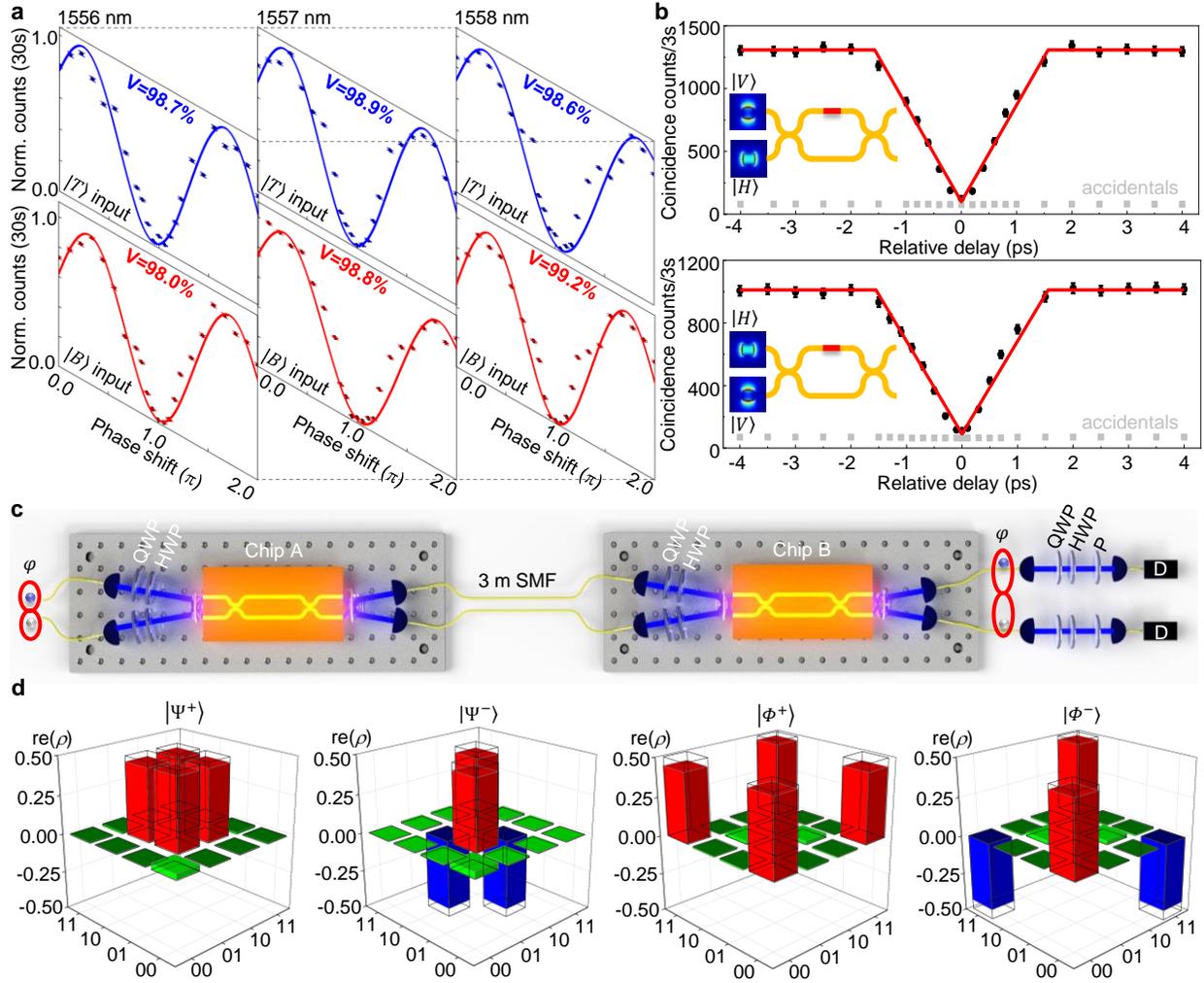

**Figure 4 | Coherence preservation of the SWAP gate operation verified via single-photon and two-photon interferences and quantum photonic interconnect between two silicon chips. a**, A polarization qubit $|H\rangle + e^{i\varphi}|V\rangle$ is sent to the SWAP gate to probe the phase coherence of the single-photon two-qubit SWAP operation. Interference fringes are obtained via tuning the phase shift φ by a pair of multi-order waveplates (illustrated in Figure 2a, III), while collecting coincidence counts between signal and idler photons (accumulated for 30 s). Measured at 1556 nm, 1557 nm, and 1558 nm, the phase interference has fringe visibilities of 98.7 ± 0.2 % (99.4%), 98.9 ± 0.2 % (99.3%), and 98.6 ± 0.2 % (98.9%) respectively for $|T\rangle$ input state before (after) background subtraction. For the $|B\rangle$ input state, the fringe visibilities are 98.0 ± 0.2% (98.5%), 98.8 ± 0.2 % (99.0%), and 99.2 ± 0.1 % (99.4%), respectively, before (after) background subtraction. **b,** Hong-Ou-Mandel interference between two photons after the SWAP operation for different input polarization combinations. 96.9 (92.4) ± 1.4% visibility is achieved for $|T_S V_S\rangle \otimes |B_I H_I\rangle$ input and 96.0 (91.0) ± 1.9% for $|T_S H_S\rangle \otimes |B_I V_I\rangle$ input after (before) background

subtraction, which proves the preservation of quantum coherence after the on-chip SWAP operation. Error bars are due to Poissonian statistics. **c,** Experimental scheme for quantum state distribution between two silicon SWAP gate chips. Polarization Bell states are prepared and fed into the first on-chip SWAP gate, the swapped spatial-momentum states are then transmitted through 3 m single-mode fiber (SMF) and coupled to the second on-chip SWAP gate, which converts the spatial-momentum states back to the polarization states. Polarization analyzers, consisting of a QWP, HWP and a polarizer, measure the polarization entangled states after the second on-chip SWAP gate for tomographic characterization. **d,** Real parts of the reconstructed density matrices of the polarization Bell states $|\Psi^+\rangle$, $|\Psi^-\rangle$, $|\Phi^+\rangle$ and $|\Phi^-\rangle$, with an averaged Bell state fidelity of 91.5 ± 0.8% after the chip-to-chip distribution, verifying the coherent reversible conversion of the SWAP operation between two silicon chips. The imaginary parts of the density matrices are negligible.